\documentclass[authoryear, 5p]{elsarticle}

\usepackage{geometry}                
\usepackage{graphicx}
\usepackage{amssymb}
\usepackage{epstopdf}
\usepackage{color}
\usepackage{amsmath}
\usepackage{bm}
\usepackage{natbib}
\usepackage{textcomp}

\begin{document}

\begin{frontmatter}

\title{Compound chondrule formation via collision of supercooled droplets}

\author{Sota Arakawa}
\ead{arakawa.s.ac@m.titech.ac.jp}
\author{Taishi Nakamoto}
\address{Department of Earth and Planetary Sciences, Tokyo Institute of Technology, 2-12-1 Ookayama, Meguro, Tokyo 152-8551, Japan}

\begin{abstract}
We present a novel model showing that compound chondrules are formed by collisions of supercooled droplets.
This model reproduces two prominent observed features of compound chondrules: the nonporphyritic texture and the size ratio between two components.
\end{abstract}


\end{frontmatter}

\sloppy

\section{Introduction}
\label{I}
Chondrules are spherical particles of $0.1$--$1\ \mathrm{mm}$ in size contained within chondrites, the most common type of meteorites, as a major component.
The volume fraction of chondrules in typical chondrites, i.e., ordinary chondrites, is up to 70\% \citep{R2000}.
It implies that the total mass of all the chondrules in all the asteroids would be very large.
This cannot be ignored compared to the total solid mass in the asteroid belt.
The ages of chondrules are about $4.6$ Gyr, slightly (probably a few Myr or less) younger than the Calcium-Aluminum-rich Inclusions (CAIs) \citep[e.g.,][]{DC2011}.
Therefore, the formation of chondrules must be related to the formation of the solar system itself, especially to the formation of the asteroids, the rocky planets, and probably Jupiter as well.
Revealing the chondrule formation process, therefore, is one of the keys to elucidating the solar system formation.
Chondrules are thought to be melted by some heating processes in the early solar nebula and become spherical due to the surface tension.
However, heating processes responsible for chondrule formation and their details remain unclear.

Studies of chondrules in thin sections have revealed that some chondrules are composed of two or more chondrules fused together.
They are called {\it compound chondrules}.
The presence of compound chondrules is interpreted to be the result of collisions among non-solid precursors.
Since the compound chondrule formation includes multiple precursors, it is a more complicated phenomenon than single chondrule formation.
It suggests that by studying compound chondrule formation, we can obtain more information on the chondrule formation process itself, since it is likely that compound chondrules and single chondrules are formed by similar mechanisms.
Determining the compound chondrule formation process is an important issue to be addressed.

Previous studies on compound chondrule formation mainly analyzed the fraction of compound chondrules among all the chondrules, which is expressed by $f_{\mathrm{compound}}^{\mathrm{ALL}}$ and is about 4\% ($f_{\mathrm{compound}}^{\mathrm{ALL}} = 4 \times 10^{-2}$) \citep{GK1981}.
Most of the previous studies assume that compound chondrules are formed by collisions of molten precursors (the molten-collision model) \citep[e.g.,][]{C2004b}.
The fraction of compound chondrules was estimated from the fraction of chondrules undergoing collisions with other chondrules \citep{GK1981,C2004b}.
The fraction of chondrules that underwent collisions with others, $f_{\mathrm{col}}$, can be estimated by
\begin{equation}
 f_{\mathrm{col}} = n \sigma v t_{\mathrm{col}},
\end{equation}
where $n$ is the number density of chondrules, $\sigma$ is the collisional cross section, $v$ is the collision velocity, and $t_{\mathrm{col}}$ is the duration of time when compound chondrule forming collisions take place.
Each quantity may be evaluated as follows.
A typical radius of chondrules $r$ is about $2 \times 10^{-2}\ \mathrm{cm}$ \citep{R2000}, and the collisional cross section of chondrules $\sigma$ is about $\sigma \sim \pi {( 2 r)}^{2} = 5 \times 10^{-3}\ \mathrm{cm}^{2}$.
We assume that the collision velocity is $v = 1 \times 10^{2}\ \mathrm{cm}\ \mathrm{s}^{-1}$ from hydromechanical constraints \citep[e.g.,][]{AP1990}.
As for the duration of time when chondrules stay molten, it is suggested to be shorter than $10^{4}$ sec based on the cooling rate estimation of chondrule formation \citep{DC2002}, or some studies suggest that the duration of time when chondrules stay molten is only a few sec \citep[e.g.,][]{YW2002}.
In order to reproduce the observed fraction of compound chondrules, $f_{\mathrm{compound}}^{\mathrm{ALL}} =0.04$, the number density $n$ should be $8 \times 10^{-6}\ \mathrm{cm}^{-3}$ or much higher.
Using a similar argument, \citet{C2004b} inferred that chondrules would have formed in regions of the solar nebula that had highly concentrated solids.
However, it is not clear if such a concentrated region can be present in the early solar nebula.
For example, when we suppose that compound chondrules are formed in the mid-plane of the minimum mass solar nebula \citep{H1981} at $2\ \mathrm{AU}$, and suppose that the dust-to-gas mass ratio is raised to be unity due to dust sedimentation, and suppose that all the dust forms chondrule precursor particles, then the estimated number density of chondrule precursor particles would only be about $2 \times 10^{-6}\ \mathrm{cm}^{-3}$ when the internal density of precursor particles is about $3\ \mathrm{g}\ \mathrm{cm}^{-3}$.
Since the values of $\sigma$ and $v_{\mathrm{c}}$ would not be changed significantly, the short duration $t_{\mathrm{col}}$ requires a higher number density $n$ to form the large fraction of compound chondrules.

Compound chondrules have some other noteworthy features.
A component in a compound chondrule usually holds spherical shape, which is called {\it primary}.
In contrast, the other component is usually deformed and called {\it secondary}.
The median size ratio of primary to secondary is approximately four \citep{W1995}; i.e., the spherical primary is typically four times larger than the deformed secondary.
In addition, compound chondrules can be grouped into three types according to structure \citep{W1995}: (1) an adhering type, wherein a small secondary is stuck on a large primary, (2) a consorting type, where both primary and secondary have roughly the same size, and (3) an enveloping type, where a secondary encloses a primary.
In ordinary chondrites, it is found that enveloping compound chondrules are rare (only eight out of 80 samples or about 10\% of all the compound chondrules) \citep[see][]{W1995}, so in this study we consider only the adhering and the consorting types.
According to \citet{W1995}, the fraction of the adhering type is $66/80 = 82.5\%$ and that of the consorting type is $6/80 = 7.5\%$.

The textures of chondrules contained in compound chondrules have another noteworthy feature.
In general, the textures of chondrules are classified into three types: porphyritic, nonporphyritic, and glassy.
According to \citet{GK1981}, only 16\% of all the chondrules are nonporphyritic chondrules, while 84\% of them are porphyritic.
Glassy chondrules are extremely rare \citep{KR1994}.
In contrast, when we look at components in compound chondrules in ordinary chondrites, we can find that most of them have nonporphyritic texture.
Although the majority of compound chondrules in CV chondrites is porphyritic, the trend ($f_{\mathrm{compound}}^{\mathrm{NP}} / f_{\mathrm{compound}}^{\mathrm{ALL}} \gg 1$) is common \citep{AN2005}, where $f_{\mathrm{compound}}^{\mathrm{NP}}$ is the fraction of compound chondrules in nonporphyritic chondrules.
\citet{W1995} revealed that 52 primaries and 65 secondaries in 72 adhering and consorting compound chondrules are nonporphyritic, so the fractions of the nonporphyritic type are $52/72 = 81\%$ for the primary and $65/72 = 90\%$ for the secondary, both of which are much higher than the fractions in all the chondrules.
Experimental studies showed that nonporphyritic chondrules are formed from completely molten droplets, and porphyritic ones are formed from partially molten particles \citep[e.g.,][]{C1998}.
Therefore, it is strongly suggested that compound chondrules are formed mainly from completely molten droplets, while single chondrules are formed from partially molten particles.

Experiments \citep[e.g.,][]{N2006,N2008} showed that completely molten levitated droplets having no contact with any other solids turned into a supercooled state at their liquidus temperature as they are cooled.
The supercooled droplets behaved as fluid particles even at a temperature lower than the liquidus.
If the temperature is maintained properly, the droplets remained supercooled for a long time.
However, when the droplets collide with a solid particle, they immediately crystallize and form a nonporphyritic texture.
Since the majority of components in compound chondrules are likely to be formed from completely molten droplets, it seems natural to expect that most of the compound chondrules have experienced a supercooled state in their formation process.
However, the supercooling has never been taken into consideration in the context of compound chondrule formation.

In this study, compound chondrule formation with supercooling is examined.
Table \ref{3features} lists some apparent features and the number of observed compound chondrules, which should be explained by a successful compound chondrule formation model.
We will address these properties, and we will see that they can be explained by taking the supercooling into account in the model.

\begin{table*}[!t]
\caption{Three features of compound chondrules that should be explained in this study.}
\label{3features}
\begin{center}
\scalebox{1.2}{
\footnotesize
\renewcommand{\arraystretch}{1.5}
\begin{tabular}{|c|c||c|} \hline
Feature I & textural type of & almost all ($90\%$) \\
 & compound chondrule & are nonporphyritic \\ \hline
Feature {I\hspace{-.1em}I} & size ratio between & 4:1 \\
 & primary to secondary & (primary is larger) \\ \hline
Fraction & compound chondrule fraction & $f_{\mathrm{compound}}^{\mathrm{NP}} = (4\% \times 90\%) / 16\% \sim 20\%$ \\
 & in nonporphyritic (or porphyritic) chondrule & ($f_{\mathrm{compound}}^{\mathrm{P}} = (4\% \times 10\%) / 84\% \sim 0.5\%$) \\ \hline
\end{tabular}
}
\end{center}
\end{table*}

\section{Crystallization of melts}
\label{C}
Understanding the crystallization mechanism is essential for discussing the formation process of compound chondrules.
However, in previous studies of compound chondrule formation, the crystallization process was considered incorrectly.
For example, \citet{H2015} assumed that the viscosity of dust particles changed continuously from the liquidus temperature to the glass transition point; however, in reality, the viscosity increases discontinuously at the solidus temperature.

\begin{figure}[!b]
\begin{center}
\includegraphics[width = \columnwidth]{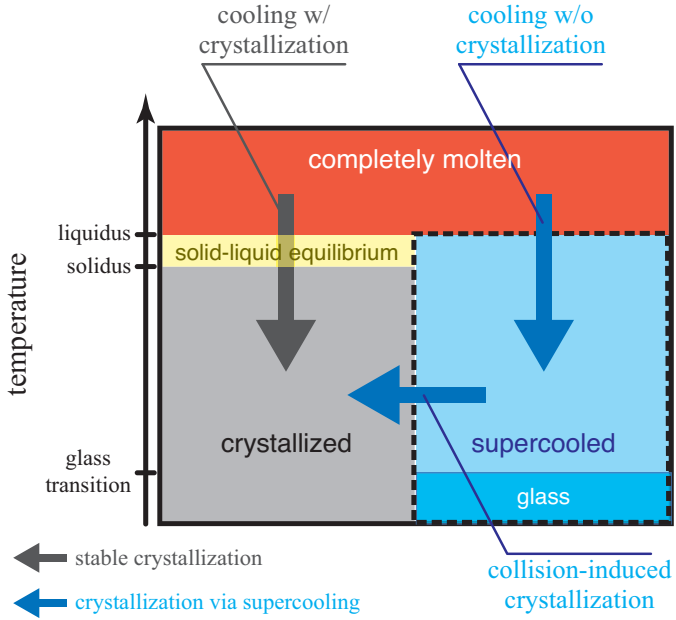}
\caption{A schematic phase diagram of dust particles in stable states (completely molten, solid-liquid equilibrium, and crystallized) and metastable states (supercooled and glass).
Previous studies considered that nonporphyritic chondrules are crystallized between their liquidus and solidus (gray arrow).
In contrast, completely molten levitated droplets invariably turn into supercooled droplets and these supercooled droplets turn into crystallized particles by contact (blue arrows).}
\label{temp}
\end{center}
\end{figure}

Figure \ref{temp} shows a schematic phase diagram of dust in stable states (completely molten, solid-liquid equilibrium, and crystallized) and metastable states (supercooled and glass).
A metastable state is an unstable equilibrium state of a macroscopic system in which the system can remain for a long period.
A supercooled droplet and a glass particle are well-known examples of metastable states.

In stable states, as the temperature decreases, completely molten droplets (colored by red in Figure \ref{temp}) turn into partially molten particles (solid-liquid equilibrium phase, yellow) at their liquidus ($\sim 1700$--$2000\ \mathrm{K}$) \citep{HR1990}, and partially molten particles turn into crystallized particles (gray) at their solidus (e.g., $1830\ \mathrm{K}$ for the $\mathrm{Mg} \mathrm{Si} \mathrm{O}_{3}$-$\mathrm{Mg}_{2} \mathrm{Si} \mathrm{O}_{4}$ system) \citep{BA1914}.
On the other hand, in metastable states, completely molten droplets turn into supercooled droplets (light blue) at the liquidus, and supercooled droplets turn into glass particles (deep blue) at the glass transition point \citep[e.g., $1063\ \mathrm{K}$ for enstatite, reviewed by][]{S1984}.

Previous studies of compound chondrule formation did not take supercooling into consideration.
Although partially molten droplets immediately turn into crystallized particles, experimental studies \citep{N2006,N2008} revealed that completely molten levitated droplets invariably turn into supercooled droplets.
\citet{N2006} studied supercooling and crystallization of forsterite droplets using levitation experiments, and they revealed that the duration of supercooling is longer than $10^{3}$ sec even if supercooled droplets are kept at low temperature ($1160\ \mathrm{K}$).
In addition, a theoretical study by \citet{T2008} suggested that the duration of supercooling can be arbitrarily long if the temperature of these supercooled droplets are kept above their glass transition points.
These supercooled droplets turn into crystallized particles by contact \citep{N2006}.
We, therefore, suggest that the crystallization of supercooled droplets is the key for the compound chondrule formation because almost all the compound chondrules have experienced complete melting.

In this study, we assume that completely molten droplets turn into supercooled droplets as their temperature decreases, and supercooled droplets turn into crystallized particles when they collide with other particles.
Even though supercooled droplets may turn into partially molten droplets, we ignore this for simplicity.
The outline of our model should hardly change with this transition.
More details of the crystallization of supercooled droplets should be studied in the future.

\section{Supercooled-collision model for compound chondrule formation}
\label{S}
We propose a novel compound chondrule formation model, wherein compound chondrules are formed through the collision and the crystallization of supercooled droplets.
We call this model a supercooled-collision model.
In this model, the heating mechanism for the chondrule formation is not specified.
Therefore, this model is a general one that could be applicable to various chondrule formation models.

\begin{figure}[!t]
\begin{center}
\includegraphics[width = \columnwidth]{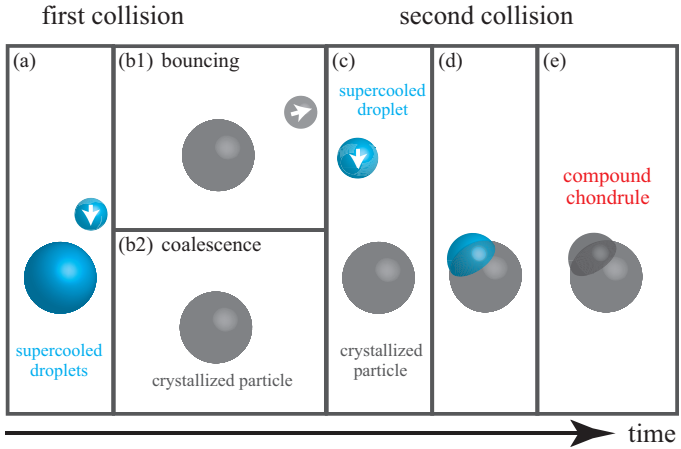}
\caption{An outline of the supercooled-collision model.
The compound chondrules are formed by collisions between supercooled droplets and crystallized particles.}
\label{outline}
\end{center}
\end{figure}

Figure \ref{outline} shows an outline of the supercooled-collision model.
We suppose that at first the temperature of the dust particles is above the liquidus, and the particles are completely molten.
Afterwards, these particles cool below the liquidus and become supercooled droplets (Figure \ref{outline}-a).
The supercooled droplets are crystallized by collisions with other particles and change into crystallized particles.
When two droplets collide with a large relative velocity, the two droplets fragment into small pieces or separate after the collision and change into two crystallized particles (Figure \ref{outline}-b1).
On the other hand, when two droplets collide with a small relative velocity, the two droplets coalesce and change into one crystallized particle (Figure \ref{outline}-b2).
When a supercooled droplet collides with a crystallized particle (Figure \ref{outline}-c), the supercooled droplet sticks to the crystallized particle (Figure \ref{outline}-d), and a compound chondrule is formed by rapid crystallization (Figure \ref{outline}-e).

The supercooled-collision model immediately reproduces two important features of compound chondrules, Feature I and Feature {I\hspace{-.1em}I} in Table \ref{3features}.
The observed Feature I---that most of the compound chondrules are nonporphyritic---can be reproduced, because a complete molten state is supposed first in the supercooled-collision model so that molten droplets can experience supercooling, and complete molten droplets can finally form nonporphyritic chondrules.
The observed Feature {I\hspace{-.1em}I}---that the primaries (spherical-shaped particles in compound chondrules) have larger sizes than the secondaries (deformed particles) in almost all the compound chondrules---can be explained by the size dependence of the collision rate.
Larger particles have a larger geometrical cross section for collision, so they collide with other particles more frequently than do the smaller particles.
In addition, it is often the case that larger particles have a greater velocity than smaller particles \citep[e.g.,][]{OC2007}, so again they tend to collide with other particles more frequently, though this depends on the chondrule forming mechanism and the environment.
Thus, larger supercooled droplets are likely to crystallize earlier than smaller droplets and become the primaries of compound chondrules.

It may be not clear what happens after stage (c) in Figure \ref{outline} for the collision of the supercooled droplet.
We need to determine if hydrodynamic motion (Figure \ref{outline}-d) or crystallization (Figure \ref{outline}-e) occurs, and we do this by evaluating the time scales of these phenomena.
Laboratory experiments suggest that the crystallization time scale is of the order of $1$ sec \citep{N2006,N2008}.
On the other hand, the time scale of the hydrodynamic motion of the supercooled fluid caused by collision should be about $r/v$, where $r$ is the radius of the droplet and $v$ is the collision velocity.
The size of the colliding droplet, which would become secondary, is of the order of $10^{-2}\ \mathrm{cm}$, and the collision velocity should be about $10^{2}\ \mathrm{cm}\ \mathrm{s}^{-1}$, as will be discussed later.
Therefore, the collision time scale would be of the order of $10^{-4}$ sec.
Since the time scale of the hydrodynamic motion is shorter than the crystallization time scale, we can see that the hydrodynamic motion takes place first and the crystallization follows.
The supercooled droplet colliding into the crystallized particle behaves as a fluid and deforms first; and after stopping the hydrodynamic motion, it crystallizes.
This estimate is consistent with the experimental results \citep{C1994}.

\section{Quantitative evaluation of the supercooled-collision model}
\label{Q}
The observed feature named {\it Fraction} in Table \ref{3features} (the fraction of compound chondrules among all the nonporphyritic chondrules, $f_{\mathrm{compound}}^{\mathrm{NP}}$) is about 20\% \citep{C2004b}.
We need to evaluate if the supercooled-collision model can quantitatively reproduce
this observed feature of the compound chondrules.

In the supercooled-collision model, the fraction of the compound chondrules among all the nonporphyritic chondrules $f_{\mathrm{compound}}^{\mathrm{NP}}$ is given by $f_{\mathrm{compound}}^{\mathrm{NP}} = n \sigma v t_{\mathrm{form}}$.
Therefore, we suppose that the compound chondrule forming collision velocities $v$ are $1 \times 10^{2}\ \mathrm{cm}\ \mathrm{s}^{-1}$, the same as the collision velocities in the molten-collision model; and we have
\begin{equation}
n t_{\mathrm{form}} = \frac{f_{\mathrm{compound}}^{\mathrm{NP}}}{\sigma v} = 4 \times 10^{-1}\ \mathrm{cm}^{-3}\ \mathrm{s}.
\end{equation}
Note that the supposed collision velocity $v$ may be different among the compound chondrule formation models, though the cross section $\sigma$ is independent of the models.

In the molten-collision model, it is supposed that the upper limit of $t_{\mathrm{form}}$ is $10^{4}$ sec (or might be a few sec) and $v$ is about $1 \times 10^{2}\ \mathrm{cm}\ \mathrm{s}^{-1}$, so the number density should be higher than $4 \times 10^{-5}\ \mathrm{cm}^{-3}$ to reproduce a sufficient amount of the compound chondrules.
However, the required number density seems to be too high to be realized in the solar nebula.
For example, as discussed in section \ref{I}, the number density of chondrule sized particles at 2 AU may reach $4 \times 10^{-4}\ \mathrm{cm}^{-3}$ only if the dust-to-gas mass ratio is about 20:1 in the mid-plane of the minimum mass solar nebula \citep{H1981}.

In contrast, in the supercooled-collision model, the duration of the supercooled state, which can be the duration of compound chondrule formation $t_{\mathrm{form}}$, can be long.
A theoretical study on nucleation \citep{T2008} indicated that the duration of the supercooled state can be arbitrarily long depending on the temperature cooling rate.
For example, it was shown that the supercooled state could last for $10^{5}$ sec or $10^{6}$ sec, when the cooling rate of the temperature is $10^{-2}\ \mathrm{K}\ \mathrm{s}^{-1}$ or $10^{-3}\ \mathrm{K}\ \mathrm{s}^{-1}$, respectively.
In fact, these slow cooling rates are considered to be appropriate as the cooling rates for chondrule formation ($\sim 10^{-3}$--$1\ \mathrm{K}\ \mathrm{s}^{-1}$) \citep{D2012}.
We think that the duration of the supercooled state can be $10^{5}$ sec or $10^{6}$ sec.

Generally, the supercooled-collision model requires a lower number density compared to the molten-collision model, since the duration for formation $t_{\mathrm{form}}$ can be long because of the supercooling.

\section{Discussion: application to chondrule formation models}
\label{D}
The supercooled-collision model for compound chondrules proposed in this study is independent of the chondrule formation models.
However, it requires certain relationships between the number density of chondrule precursors $n$ and the duration of compound chondrule formation $t_{\mathrm{form}}$ as discussed in section \ref{Q}.
We now examine if some chondrule formation models studied to date may meet the requirement for compound chondrule formation.
The examined chondrule formation models include the planetesimal bow shock model, the lightning model, and the impact jetting model.

The planetesimal bow shock model assumes that some eccentric planetesimals in the solar nebula generate bow shocks around them; and the chondrule precursor particles in the solar nebula are heated by the bow shock, mainly due to the gas frictional heating \citep[e.g.,][]{M2012,N2014}.
According to \citet{M2012}, chondrules heated in the shocked gas are cooled to the glass transition point about 10 hours ($\simeq 4 \times 10^{4}$ sec) after heating.
This time scale can be regarded as the duration of the supercooled state and the duration of compound chondrule formation $t_{\mathrm{form}}$.
We then obtain that the required number density of the chondrule precursors is about $1 \times 10^{-5}\ \mathrm{cm}^{-3}$, which might be realized in a dust-rich region such as the dust-layer at the mid-plane of the solar nebula.
Therefore, the planetesimal bow shock model may be consistent with compound chondrule formation using the supercooled-collision model.

The planetesimal bow shock model may provide another benefit for the compound chondrule formation.
\citet{T2008} showed that supercooled forsterite droplets are crystallized without collision near the transition temperature ($\sim 1200\ \mathrm{K}$) and that they do not turn into glass.
Forsterite-rich particles crystallize spontaneously.
In contrast, enstatite-rich particles never crystallize spontaneously and turn into glass at the transition temperature.
To crystallize enstatite-rich particles, some solid particles as nuclei for crystallization are needed \citep{N2008}.
Since natural chondrites do not contain many glass chondrules, some nuclei for crystallization should be supplied for enstatite-rich particles in the chondrule forming.
These nuclei may be provided as tiny solid particles, which are condensates from silicate vapor in the chondrule forming region.
\citet{Miura-et-al-2010} suggested that nm- and {\textmu}m-sized dust grains are formed in the shocked region, at the temperature around $1000\ \mathrm{K}$, which is near the glass transition point.
Because of the supply of these small grains, enstatite-rich glassy chondrules are not formed frequently.

The lightning model \citep[e.g.,][]{H1995} assumes that charged dust particles in the solar nebula are decoupled with gas, and the motions of the dust particles generate local potential differences in the solar nebula.
Electrical charges are released then as lightning.
There are several positive \citep[e.g.,][]{M2010} and negative \citep[e.g.,][]{G2008} opinions regarding this mechanism.
However, from the viewpoint of compound chondrule formation, the lightning model seems to be inappropriate.
\citet{H1995} suggests that the duration of melting is extremely short (the temperature of the dust particles decreases at the glass transition point within 10 sec).
In this case, the fraction of compound chondrules cannot be reproduced by the lightning model.

The impact jetting model \citep[e.g.,][]{J2015} assumes that chondrules are formed by collisions between planetesimals or protoplanets.
According to \citet{J2015}, the dust particles ejected by the collisions are mm-sized completely molten droplets, and cooling rates of these particles are of the order of $10^{-1}\ \mathrm{K}\ \mathrm{s}^{-1}$.
These features are consistent with the constraints of chondrule formation.
However, the number of collisions per dust particle calculated by \citet{J2015} is of the order of $10^{2}$, which might be extremely large for reproducing the fraction of compound chondrules.
This high frequency of collision leads to the conclusion that not only nonporphyritic chondrules, but also most porphyritic ones, are compound.
In reality, the fraction of compound chondrules in porphyritic chondrules is only $0.5\%$.
Therefore, the impact jetting model might not be appropriate at least for the compound chondrules in ordinary chondrites.

\section{Conclusion}
\label{C}
We examined if the supercooled-collision model, a novel model for compound chondrule formation schematically shown in Figure \ref{outline}, can reproduce the observed features of compound chondrules in ordinary chondrites.
Compound chondrules have three features: there are nonporphyritic textural types (Feature I), the primary is about four times as large as the secondary (Feature {I\hspace{-.1em}I}), and $20\%$ of the nonporphyritic chondrules are compound (Fraction).
Feature I suggests that almost all of the compound chondrules have experienced complete melting \citep{C1998} and supercooling \citep{N2006}.
Feature {I\hspace{-.1em}I} can be explained by the supercooling and crystallization of dust particles.
Large supercooled droplets tend to collide with other particles frequently compared to small droplets, with large particles tending to turn into the primaries and the small supercooled survivors sticking on the primaries.
The fraction of compound chondrule is related to the number density of the dust particles and the duration of supercooling.
Because of the long duration of supercooling \citep{T2008}, the required number density can be low.
Therefore, we conclude that compound chondrules are likely to be formed through collisions among supercooled droplets.

\section*{Acknowledgments}
We thank Kyoko K.\ Tanaka and Katsuo Tsukamoto for their valuable discussions based on their invaluable data, and Harold C.\ Connolly Jr., Hitoshi Miura, Makiko Nagasawa, Ken Nagashima, Tomoki Nakamura, Hidekazu Tanaka, and the anonymous reviewers for their helpful comments.
This work is supported by JSPS KAKENHI Grant Number 15K05266.


\bibliographystyle{elsarticle-harv}

\bibliography{compound}

\end{document}